# Experiments on transformation microfluidics: cloaking flow and transport without metamaterials

Oscar Boyadjian[1], Etienne Boulais[2] and Thomas Gervais[2, *]

*[1]Biomedical Engineering Institute, École Polytechnique de Montréal, Montréal, QC, H3T 1J4 Canada*
*[2]Engineering Physics Department, École Polytechnique de Montréal, Montréal, QC, H3T 1J4 Canada*

(Dated: December 15, 2020)

Cloaking effects have now been identified in almost every field of physics. In all cases, substrate-carved metamaterials make the reshaping of the concealed volume impossible. In fluids, recognizing that d'Alembert's paradox describes in itself a hydrodynamic cloaking mechanism, we propose and characterize experimentally a framework to cloak both flow and transport simultaneously within a reshapable fluid volume. The properties of microfluidic cloaks as open-space diffusion filters (O-Filters) are further investigated. They reveal a strong topological analogy with classical H-Filters in microfluidics.

## I. INTRODUCTION

In their 2006 seminal papers, U. Leonhardt and J.B. Pendry et al independently introduced a way to redirect light around an obstacle, laying the foundations for optical cloaking [1][2] and opening up the field of transformation optics [3][4]. Their mathematical framework was found to be quite general and applicable to a variety of physical problems [5]. This has led to the development of a host of cloaking devices in other fields of physics by using metamaterials to encode conformal maps of a physical parameter into a medium. For instance, cloaking waveguides have been engineered for acoustic waves [6][7], elastic waves [8][9], seismic waves [10], fluid surface waves [11] and even quantum mechanical matter waves [12].

More recently, cloaking has also been achieved for physics not bound by parabolic wave equations but still underlain by conformally invariant equations, such as heat diffusion in solid bodies [13][14] and flow in porous media [15]. Even though some metamaterials may exhibit thermally [16], mechanically [17] or optically [18] tunable cloaking properties thanks to nonlinear components, the concealed area remains fixed. This limitation comes inherently from substrate-carved metamaterials. Similarly, non-reshapable flow cloaks also exist in fluids since J. Park et al. introduced a way to conceal a solid cylinder from an incoming Stokes flow by using microstructured chambers to yield a drag-vanishing viscous tensor [19].

Active reshaping of the cloaking volume to successively cover or uncover an object or a surface would be extremely useful in many applications. In microfluidics, for example, experiments usually employ sensitive surfaces such as cell cultures [20], patterned antibodies [21], or surface-based biosensors [22] that need to be selectively protected from a chemically or thermally hostile environment until they must be exposed to it for sensing. To our knowledge, no such double cloaking mechanism, concealing both flow and diffusive transport simultaneously, has ever been proposed in the literature. It is however possible, in principles, in the situation where underpinning equations are naturally conformally invariant.

It turns out this is the case for both flow and transport equations in quasi 2D purely viscous flows, also called Hele-Shaw flows [23], which can easily be generated in microfluidic systems [24]. Based on mathematical advances in the conformal mapping of non-harmonic equations in transport theory made by M.Z. Bazant et al [25], our group has recently demonstrated theoretically and experimentally how conformal transforms can be used to study diffusive transport in arbitrary arrangements of sources and sinks [26]. In this letter, we use these strategies to design a metamaterial-less flow and advection-diffusion cloaking device in microfluidic Hele-Shaw cells, thus introducing the concept of transformation microfluidics. We then fabricate such cloaking device using 3D printing and experimentally prove its mechanical and chemical operation. Results show that 2D transport cloaks using this approach are open-space topological analogs of the well-known H-filter in microfluidics, hinting at applications in diffusion-based separation in open 2D flows.

## II. FLOW AND TRANSPORT CLOAKING

### A. Flow

In classical fluid mechanics, the interaction of a plane flow with several aligned mass-balanced injecting and aspirating point sources generates an exclusion zone in a flow stream called a Rankine body. They have been studied extensively to minimize the drag in ship hulls [27] or to model groundwater heat pumps [28][29]. Another way to generate them relies on electro-osmotic flows created by fixed surface charges patterns [30]. The simplest Rankine bodies are obtained by aligning a pair of mass-balanced injection and aspiration apertures, or doublet, in line with an incident stream [FIG. 1. (a,b)]. We have fabricated a device able to generate such a microfluidic Rankine body using 4 tubes connected to syringe pumps, a glass slide and a 3D-printed central cell being $W = 1.6$ cm wide, $L = 2$ cm long and $G = 500\ \mu\text{m}$ deep. This flow chamber is pierced in its center of two $a = 135\ \mu\text{m}$ diameter apertures being $d = 1$ mm apart from each other [FIG. 1. (c)]. The various parts of this device are sealed together with UV-sensitive glue, and dilute fluorescent beads ($3\ \mu\text{m}$ diameter) and dyes (FITC/TRITC) in water-ethanol mixture are injected in the device to enable flow and transport visualization, respectively [FIG. 1. (b,d)]. The device thus forms a Hele-Shaw cell of depth $G \ll (W, L)$ in which a central recirculation zone circling the apertures forms within the lateral flow. We see that diffusion has a significant influence on the concentration map, especially downstream [FIG. 1. (d)].

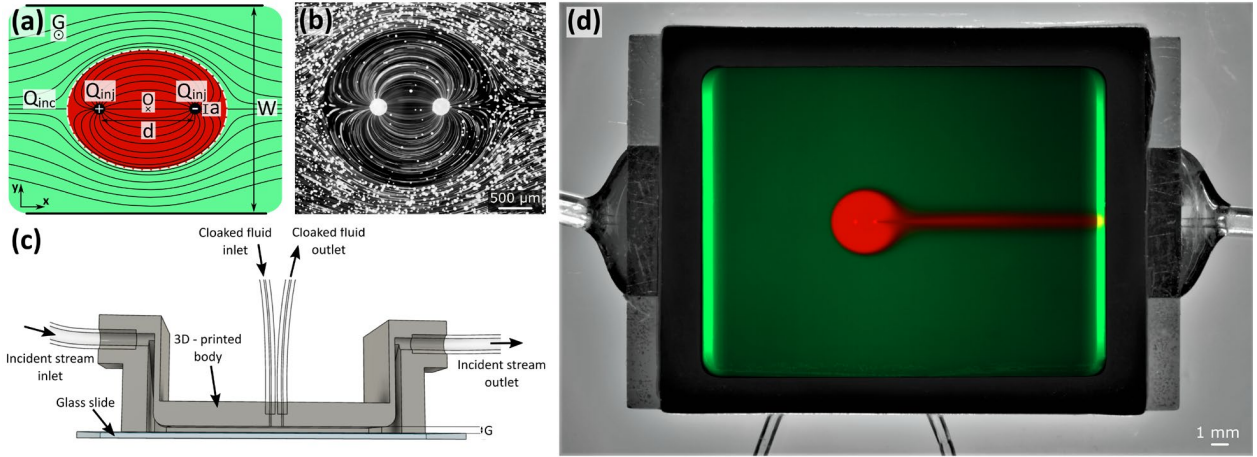

FIG. 1. Microfluidic cloak paradigm. (a) Annotated schematic representation of cloak's flow. (b) Experimental picture of cloak's flow streamlines (c) Cutaway CAD drawing of the experimental device. (d) Picture of an operating cloak visualized by injecting fluorescein in the incident stream (green) and rhodamine B in the recirculation zone (red). Black strips may be due to the Förster resonance energy transfer (FRET) occurring between these two fluorophores [P. Zhao and S. Rodriguez, Journal of Fluorescence 7, 121 (1997)].

The effects of the lateral walls over velocity being negligible in our experiments (see Supplemental Material [31]), the complex flow potential for our 2-aperture configuration can be expressed in terms of a single complex variable $z = x + iy$ by superposing two-point sources and an incident stream in a Hele-Shaw cell:

$$\Phi(z) = \frac{Q_{inc}}{WG} z + \frac{Q_{inj}}{2\pi G} \left( \log\left( z + \frac{d}{2} \right) - \log\left( z - \frac{d}{2} \right) \right), \quad (1)$$

Where $Q_{inc}$ and $Q_{inj}$ are the flow rates of the incident stream and at the apertures, respectively. With flow rates magnitude on the order of $1\ \mu L/s$ for $Q_{inc}$ and $Q_{inj}$, the chamber's Reynolds number is calculated to be ~0.001, which ensures a viscous-driven flow regime governed by Stokes' equation.

Making Eq. (1) dimensionless yields

$$\tilde{\Phi}(\tilde{z}) = \beta \tilde{z} + \log(\tilde{z} + 1) - \log(\tilde{z} - 1), \quad (2)$$

in which the dimensionless velocity ratio $\beta = \pi d Q_{inc} / WQ_{inj}$ appears and dictates alone the size of the Rankine body. To simplify the notation from here on, we use $\Phi$ to refer to the dimensionless potential [Eq. (2)]. In such configuration, a closed ellipsoidal recirculation zone isolated from the rest of the stream appears. In the particular case of an infinite 2D medium, the interface between two fluids undergoes a slip boundary condition, and the obstacle generates no drag, as per d'Alembert's paradox [32]. Simple scaling laws further reveal that when the medium is bound by lateral walls, the drag only rises as the square of the blockage ratio. Such second order correction is negligible in typical experimental situations (see Supplemental Material [31]). This recirculation zone is thus a velocity and pressure cloak as flow variables are not disturbed by it afar.

### B. Transport

The concentration profile in this microfluidic cloak is in turn determined by the 2D steady-state convection-diffusion equation:

$$\Delta c + Pe\nabla\Phi \cdot \nabla c = 0. \quad (3)$$

As it turns out, Eq. (3) in the context of 2D Hele-Shaw microfluidics is one of the handful of equations beyond Laplace's equation which are conformally invariant [25]. By using transformation to streamline coordinates $(\varphi, \psi) = (\text{Re}(\Phi), \text{Im}(\Phi))$, the problem becomes that of two incoming plane flows separated by a horizontal obstacle (see Supplemental Material [31]).

At high Peclet numbers, the problem becomes similar to the problem of convection-diffusion around a finite absorber, for which no simple analytical solution exists [33]. We can however obtain a concise, closed-form solution for high Peclet number flows that is valid near the leading edge of the cloak.

The solution [25][34], given by

$$c(\varphi,\psi) = \begin{cases} \dfrac{1}{2}\left(1 - erf\left(Im\left(\sqrt{Pe}\Phi\right)\right)\right), \ \psi < 0 \\ \dfrac{1}{2}\left(1 + erf\left(Im\left(\sqrt{Pe}\Phi\right)\right)\right), \ \psi \geq 0 \end{cases}, \quad (4)$$

indicates that the concentration profile is only dependent on one dimensionless number, the Peclet number $Pe = Q_{inj}/2\pi GD$ (where $D$ denotes either thermal or mass diffusivity). We see at once that, provided the chamber is large enough, only two parameters are needed to fully describe the cloak's profile: the velocity ratio $\beta$ to account for fluid flow effects and the Peclet number for diffusion. These two parameters can be varied independently to decouple entirely cloaking shape and diffusion losses from the cloak. The cloak configuration can be modified by varying both $\beta$ and $\overline{Pe}$, as shown in FIG. 2. (a)(i)-(ix). Following [35], the images were sorted according to the local Peclet number

$$\overline{Pe} = Pe\,\beta^2\sqrt{1 + \frac{2}{\beta}}, \quad (5)$$

which ensures comparable diffusion length for images of similar $\overline{Pe}$. From Eq. (1), the cloak's horizontal "major axis" and vertical "minor axis" can also be extracted and show good agreement with experimental values experiments [FIG. 2. (b), Supplemental Material [31]].

As shown in FIG. 1. (d) and FIG. 2. (a), the transport aspect of cloaking is always limited by a diffusion boundary layer around the Rankine body. This uncloaked imperfection can be made arbitrarily small by increasing the Peclet number while maintaining $\beta$ constant [FIG. 2. (a), columns]. This is reminiscent of a similar property in optical cloaks, where the obstacle is never rendered totally invisible but actually made arbitrarily thin, in such a way that its surface is imperceptible by geometrical optics [1]. In this sense, the flow and transport cloaking framework introduced here constitutes the basis of what we may call "transformation microfluidics", by analogy with transformation thermodynamics [13] or transformation optics [1][2].

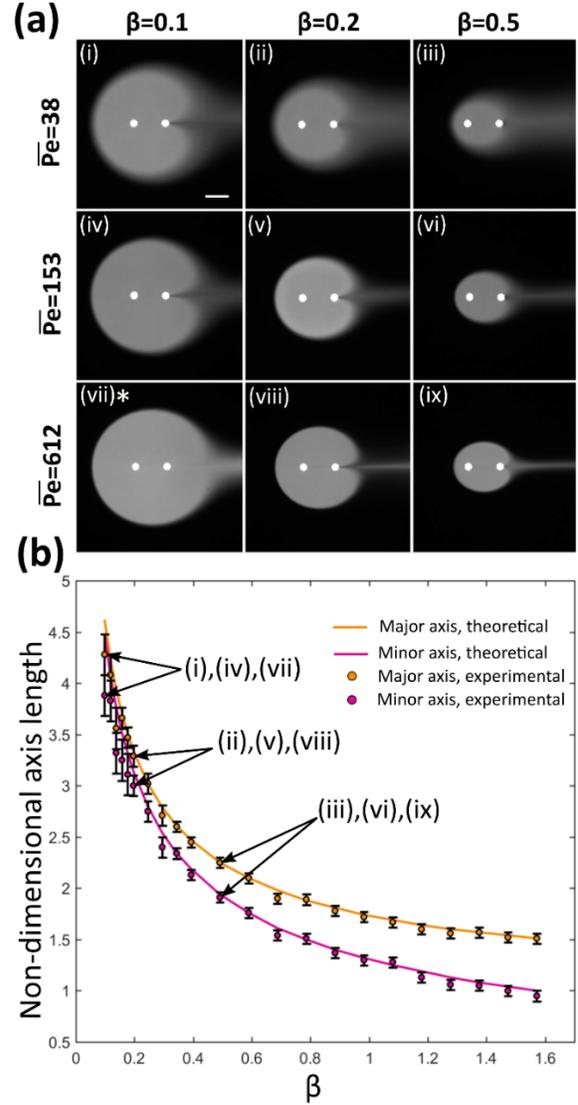

FIG. 2. Cloak control. (a) (i)-(ix) Mosaic of experimental results displaying the various cloak shapes, depending on $\beta$ and $\overline{Pe}$. Scale bar length corresponds to 1 mm. * This configuration got distorted by a lack of luer-lock airtightness during the experiment, yielding an actual $\overline{Pe}$ of 367. (b) Cloak's major and minor axis as functions of $\beta$. Theoretical curves were calculated with Eq. (2), see Supplemental Material [31]. Error bars indicate the estimated uncertainty due to the concentration gradient blurring the edge.

## III. TIME-DEPENDENT CLOAKING

Practically, the main interest in generating devices that can isolate zones from incoming flow lies in the ability to modify the exposure of a surface to outer flow, such that parts of a sensitive area can be exposed at times and protected from it at others. This drove us to consider the transient aspects of transport in microfluidic cloaks. To do so, we have studied experimentally the transition within the cloak from one steady-state to another following an abrupt change in the dimensionless velocity ratio $\beta$ [FIG. 3. (a)]. We observed that the cloak's shape transient time is governed by the convective time scale, which is found to be simply $t_c \sim L/U = 15\,s$, where $L$ is the cloaked region semi-major axis and $U$ the characteristic velocity [FIG. 3. (b)]. Meanwhile, the diffusive characteristic time is here on the order of $t_d \sim L^2/D = 1.6\,\text{h}$. The large difference between these time scales ensures an efficient shielding at high Peclet numbers, as the chemical exchanges between the cloak and its surroundings are prevented. Alternatively, it allows the delivery of downstream reagent pulses of diverse shapes by varying $\beta$ [FIG. 3. (a)]. Overall, the stability of the steady states over time without fluctuations between transitions has also been demonstrated, allowing us to protect a surface from undesired adsorption [FIG. 3. (c)].

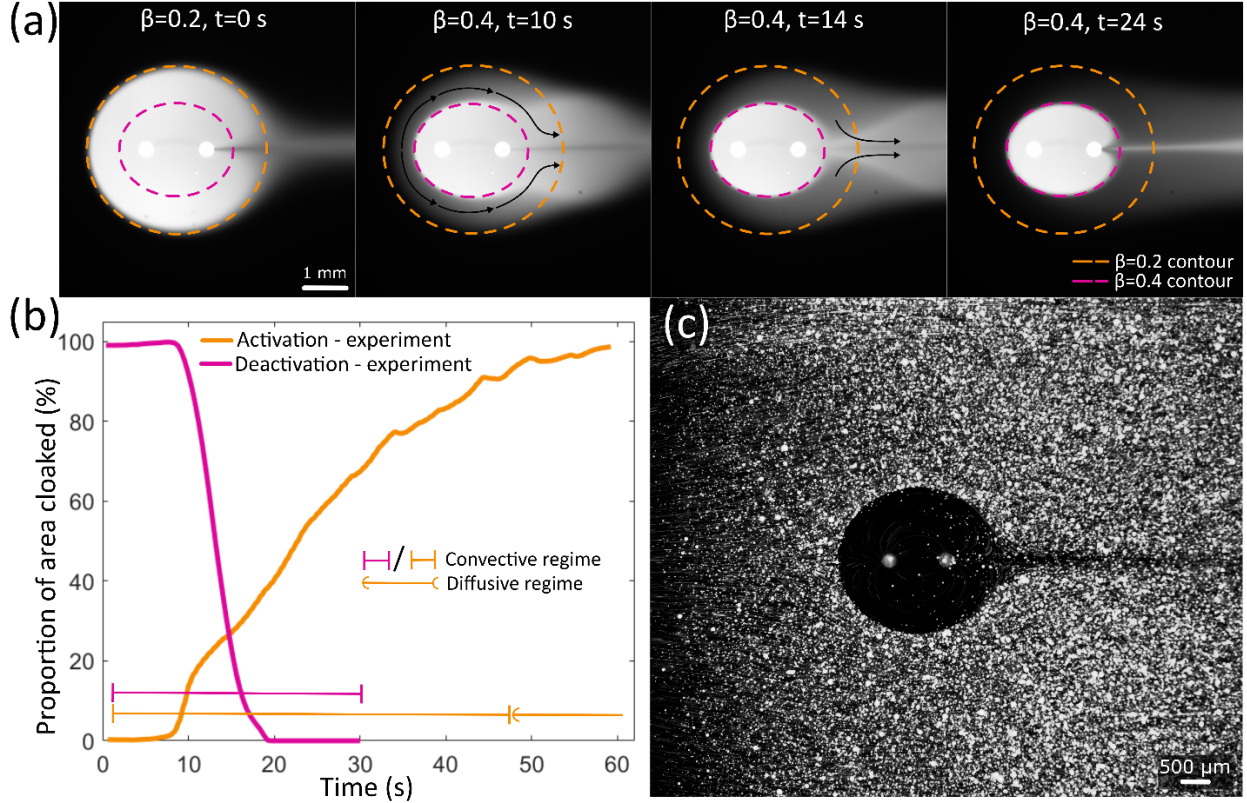

FIG. 3. Time-dependent cloaking. (a) Experimental cloak response when $\beta$ is a step function suddenly varying from 0.2 to 0.4 (b) Activation and deactivation kinetics of a cloak. (c) Experimental picture of an area protected from a fluorescent beads stream for 38 minutes. The wide luminous dots show adsorbed beads bundles.

## IV. FILTRATION: THE O-FILTER VS THE H-FILTER

The concentration profiles obtained from Eq. (4) yield outstanding agreement with experimental measurements up to very close to the devices diffusive tail at moderate to high Pe [FIG. 4. (a,b)]. Since Eq. (4) is a solution of the conformally invariant Eq. (3), it ensues that this match would be maintained under similar conditions in any other device geometry obtained via a conformal transform of the current one [36]. In this manner, by transforming the cloak geometry to streamlines coordinates, and then using Schwarz-Christoffel mapping [5], we find that the microfluidic cloak is a conformal analog of the H-Filter [FIG. 4. (c)], one of the most recognized filter in the field of microfluidics [37]. The close analogy between H-Filter and the microfluidic cloak suggests the latter can also be used as a new type of diffusion-based Open-space Filter or "O-Filter". Inspection of the dimensionless diffusive flux (or Sherwood number) around the Rankine body reveals that it scales as $\sqrt{Pe}$ [FIG. 4. (d)], as would be expected for conformally analog filtration processes arising both from transport at a slip boundary condition [38]. Contrary to H-Filters, O-Filters cover finite circular space in a chamber. Multiple individual cloaks can be placed one behind another in a Hele-Shaw cell can also be turned on or off independently in a cascade without affecting the overall flow balance in the device.

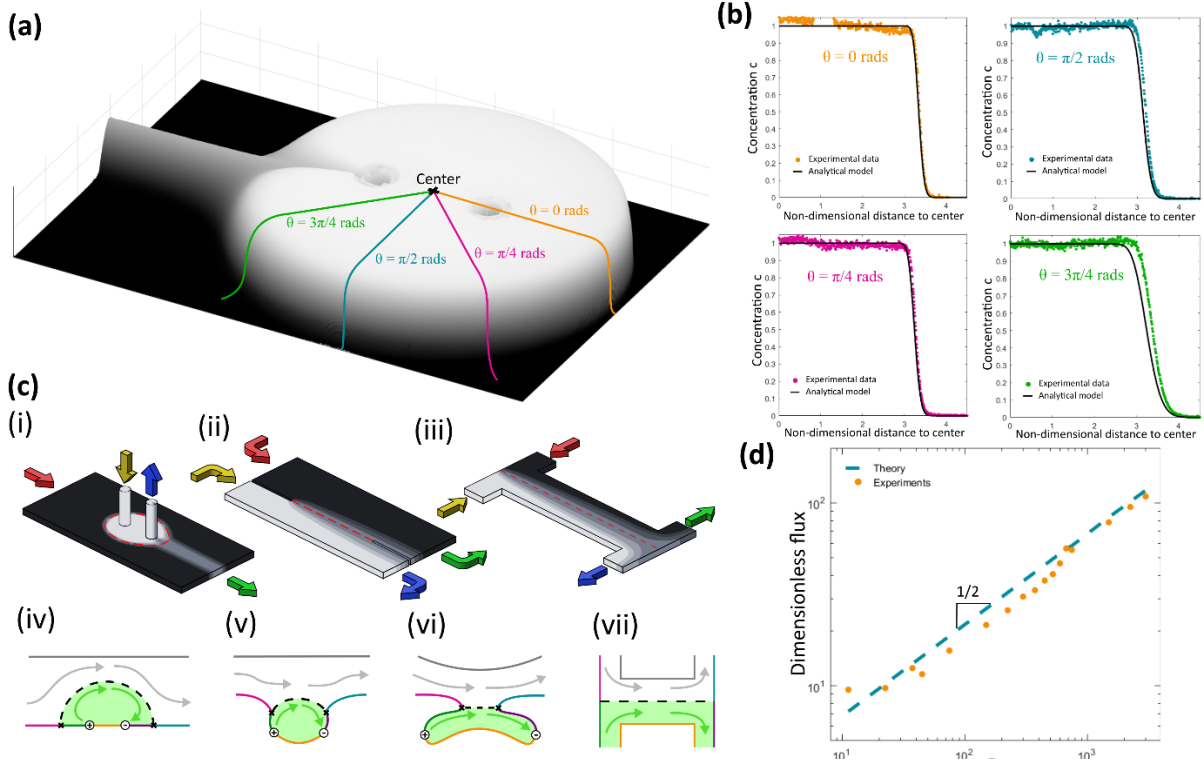

FIG. 4. Convection-diffusion within the cloak. (a) Visualizing an experimental cloak's concentration map as a surface plot (height is proportional to concentration). (b) Comparison between experimental concentration map and the model obtained with Eq. (4) at various angles. (c) Drawings of (i) cloak domain, (ii) streamlines coordinates domain, and (iii) H-filter domain. (iv)-(vii) 2D drawings of the steps to conformally transform a microfluidic cloak into a H-filter. (d) Power law between dimensionless flux and Peclet number.

## V. CONCLUDING REMARKS

Microfluidic cloaks could yield applications in fields where minimally disturbing surface reconfigurable shielding is desired, in surface-based sensing technologies like SPR sensors [39], whispering gallery mode sensors [40], quartz crystal microbalance sensing [41] and, in general, any assay where a target flows over a capture surface [22]. Given the strong mathematical analogy between mass and heat transport, they could also be used to alter temperature in a specific region of a device while leaving the surroundings unaffected or to dissipate heat locally such as in chip cooling applications [42]. The close analogy between the microfluidic cloak and the H-Filter also suggests that many more equivalences could be made between open-space microfluidics [43] and classical closed-channel microfluidics. In general, applications of conformal transforms in order to tailor transport properties in microfluidic devices is a quite recent practice pioneered by Randall et al. [44], which beckons further attention.

## ACKNOWLEDGEMENTS

This work was supported by a Fonds de Recherche du Québec − Nature et Technologie Team Grant (FRQNT #205993 2018). The authors also wish to acknowledge support from the National Science and Engineering Council of Canada (NSERC) for a Graham-Bell fellowship (E.B.) and a Discovery Grant (RGPIN − 2020 − 06838).

**REFERENCES**

* Correspondence should be addressed to thomas.gervais@polymtl.ca